\documentclass[a4paper]{article}
\usepackage{graphicx}
\usepackage{fullpage}
\usepackage{physics}
\usepackage[backend=bibtex, style=chem-acs]{biblatex}
\usepackage{amsmath}
\usepackage{braket}
\usepackage{color}
\usepackage{authblk}
\usepackage{comment}
\usepackage{bm}
\usepackage{soul}
\usepackage{xspace}
\usepackage{booktabs}
\usepackage{csvsimple}
\usepackage{url}
\usepackage{multirow}

\def\HTC {\hat{H}_{\rm TC}}
\def\e {{\rm e}}

\newcommand{\avxz}[1]{aug-cc-pV{#1}Z}

\def\avdz {\avxz{D}\xspace}

\def\avtz {\avxz{T}\xspace}

\newcommand{\neci}{\texttt{NECI}\xspace}
\newcommand{\molpro}{\texttt{Molpro}\xspace}
\newcommand{\casino}{\texttt{CASINO}\xspace}
\newcommand{\pyscf}{\texttt{pyscf}\xspace}

\newcommand{\pytchint}{\texttt{pytchint}\xspace}
\newcommand{\elemcojl}{\texttt{ElemCo.jl}\xspace}

\title{Modular Construction of Jastrow Factors for the Transcorrelated Method}
\author[1]{J. Philip Haupt\thanks{\protect\url{jphaupt@proton.me}}}
\author[2]{Maria-Andreea Filip}
\author[1]{Evelin Martine Corvid Christlmaier}
\author[1]{Yifan Cheng}
\author[1]{Johannes Hauskrecht}
\author[1,2]{Ali Alavi\thanks{\protect\url{a.alavi@fkf.mpg.de}}}
\affil[1]{Max Planck Institute for Solid State Research, Heisenbergstr 1, 70569 Stuttgart, Germany}
\affil[2]{Yusuf Hamied Department of Chemistry, University of Cambridge, Lensfield Road, Cambridge CB2 1EW, United Kingdom}

\date{}

\begin{document}

\maketitle
\begin{abstract}
In this work, we explore the reuse of terms in the Jastrow factor between systems for use in the transcorrelated method, to reduce the number of optimisable parameters for a given system. In particular, we propose a workflow in which atom-specific parts of Jastrow factors, optimised in atoms, may be reused in the molecule, with only a few parameters in the electron-electron part of the Jastrow left to optimise, while maintaining performance. We find that the modified workflow not only reduces the number of terms needing to be optimised, but also improves the accuracy of xTC-CCSD(T) energies.
\end{abstract}

\section{Introduction}

A central theme of quantum chemistry is the accurate and efficient calculation of many-body correlation. Conventional quantum chemistry methods involve expanding the wave function in terms of single-particle basis functions. However, due to cusps in the exact wave function \cite{katoEigenfunctions1957}, the convergence with respect to these basis functions is notoriously slow. Consequently, so-called ``explicitly correlated methods'' have been developed to handle this by directly addressing the cusp conditions, typically by explicit inclusion of the distance $r_{12}$ into the wave function.\cite{slaterCentral1928,hylleraasNeue1929} Of these explicitly correlated methods, F12/R12 methods have been particularly successful.\cite{kutzelnigg12Dependent1985,kongExplicitly2011}

Here, we study another explicitly correlated method, called the transcorrelated (TC) method, first developed by Boys and Handy \cite{handyEnergies1969,boysCalculation1969}. In this method, the electronic Schr\"odinger equation is similarity-transformed using a Jastrow factor\cite{jastrowManyBody1955}, which explicitly handles cusp conditions, while also introducing additional complications such as three-body excitations and non-Hermiticity. This method has recently enjoyed renewed attention\cite{ten-noFeasible2000,umezawaTranscorrelated2003,umezawaGroundstate2004,umezawaExcited2004,umezawaPractical2005,tsuneyukiTranscorrelated2008,ochiEfficient2012,luoCombining2018,cohenSimilarity2019,baiardiTranscorrelated2020,ammarExtension2022,ginerNew2021,dobrautzPerformance2022,liaoDensity2023,leeStudies2023,ammarTranscorrelated2023,dobrautzInitio2024,szenesStriking2024,ammarCompactification2024,magnussonEfficient2024,sokolovOrders2023,hauptDissertation2025,feniouRealSpace2025,hauptMultireferenceTC2025}, being combined with classical post-Hartree-Fock quantum chemistry methods, as well as quantum computing. Among the important developments are the introduction of flexible Jastrow factors and how to optimise them in the context of TC \cite{lopezriosFramework2012,hauptOptimizing2023}, and the xTC approximation, which effectively removes the three-body excitations \cite{christlmaierXTC2023}.

One of the remaining drawbacks of this method is its complexity, particularly when using highly flexible, multi-parameter, Jastrow factors. In this work, we begin to address this by reusing terms in the Jastrow factor from previously-calculated systems. For example, when using these detailed Jastrow factor forms, used in previous studies\cite{hauptOptimizing2023,christlmaierXTC2023,simulaEcps2025,filipDeterministic2025,filip_2ndrow,hauptMultireferenceTC2025} and discussed here further, the CN molecule has 74 parameters to be optimised in the Jastrow factor. In this work, we reduce the number of parameters to 8 while maintaining and even improving the accuracy.

\section{Theory}

\subsection{Transcorrelation}
Here we recapitulate the theory of the TC method. We begin with the Slater-Jastrow ansatz,
\begin{equation}
    \label{eq:slater-jastrow}
    \Psi=\mathrm e^J\Phi
\end{equation}
where $J$ is a real, symmetric function depending on the coordinates of electrons and nuclei in the system. $\Phi$ is an antisymmetric wave function for which we wish to solve. We expect $\Phi$ to be easier to solve for when $J$ is chosen to incorporate appropriate physical conditions, most notably cusp conditions, but also dynamical correlation effects. Determining $J$ has been the subject of previous work\cite{hauptOptimizing2023,hauptMultireferenceTC2025}, and here we aim to simplify this method by reducing the number of optimisable parameters in $J$.

Inserting equation \ref{eq:slater-jastrow} into the Schr\"odinger equation and premultiplying by $\e^{-J}$ results in the TC eigenvalue problem,
\begin{equation}
    \hat H_\mathrm{TC} \Phi = E\Phi
\end{equation}
where $\hat H_\mathrm{TC}=\e^{-J}\hat H\e^{J}$ is the transcorrelated Hamiltonian operator. This may be expanded by the Baker-Campbell-Hausdorff formula, which in this case truncates exactly at second order,
\begin{align}
   &\HTC = \e^{-J}\hat{H}\e^{J} = \hat{H} + [\hat{H},J] + \frac{1}{2}[[\hat{H},J],J] \nonumber \\
    &= \hat{H}
    - \sum_{i}\left(\frac{1}{2}\nabla_{i}^{2}J
              + (\nabla_{i}J) \cdot \nabla_{i}
              + \frac{1}{2}(\nabla_{i}J)^{2}\right).
    \label{eq:bch}
\end{align}
This may further be written as
\begin{equation}
    \HTC = \hat{H} - \sum_i \hat M(\bm r_{i}) - \sum_{i<j} \hat{K}(\bm r_i, \bm r_j) - \sum_{i<j<k} L(\bm r_i, \bm r_j, \bm r_k), \label{eq:htc}
\end{equation}
where $\hat M$ and $\hat K$ are a non-Hermitian one- and two-body operators respectively, and $\hat L$ is a Hermitian three-body operator.

$\hat L$ is the most computationally demanding term. However, effective approximation schemes have been developed, notably xTC,\cite{christlmaierXTC2023} which effectively reduces $\HTC$ to be a two-body operator. Within xTC, the Hamiltonian is normal ordered with respect to $\Phi$,\cite{kutzelniggNormal1997,kutzelniggCumulant1999}
\begin{equation}
 \begin{aligned}
    H_N &= \HTC - \langle\Phi|\HTC|  \Phi \rangle \\
        &= F_N + V_N + L_N,
\end{aligned}
\end{equation}
where the one-, two-, and three-body operators are (using Einstein summation)
\begin{subequations}
    \begin{align}
        F_N =
        \bigg[&
            h_P^Q
            +\big(U_{PR}^{QS}-U_{PR}^{SQ}\big)\gamma^R_S \nonumber\\
            &-\frac{1}{2}\big(L_{PRT}^{QSU}-L_{PRT}^{SQU}-L_{PRT}^{USQ}\big)\gamma^{RT}_{SU}
        \bigg]\tilde a^{P}_{Q},
    \end{align}
    \begin{align}
        V_N =\frac{1}{2} \bigg[
            U_{PR}^{QS}
            -\big(L_{PRT}^{QSU}
            -L_{PRT}^{QUS}-L_{PRT}^{USQ}\big)\gamma^T_U
        \bigg]\tilde a^{PR}_{QS},
    \end{align}
    \begin{eqnarray}
        \label{eq:LN}
        L_N =-\frac{1}{6} L_{PRT}^{QSU}\tilde a^{PRT}_{QSU},
    \end{eqnarray}
\end{subequations}
with $U=V-K$, $\tilde a^{P\dots}_{Q\dots}$ being the normal-ordered
excitation operators, and $\gamma^{P\dots}_{Q\dots}=\langle \Phi| a^{P\dots}_{Q\dots}|\Phi\rangle $ being the density matrices. The correction to the zero-body term $E_\mathrm{nuc}$ is then the expectation value of the three-body operator,
\begin{equation}
    \bra{\Phi}L\ket{\Phi} = -\frac 16L_{PRT}^{QSU}\gamma_{QSU}^{PRT}.
\end{equation}
The xTC approximation is used throughout this paper. Note that these corrections that depend on $\hat L$ may be computed without requiring a full calculation of $\hat L$ and storing its values.

\subsection{Jastrow Factors}

In the following sections, we consider various approaches to building Jastrow factors for the transcorrelated method. We use the abbreviations e-e, e-n, and e-e-n to refer to electron-electron, electron-nucleus, and electron-electron-nucleus terms in the Jastrow factors, respectively.

\subsubsection{Flexible Jastrow Factors}
\label{sec:flexible-jastrows}

In a TC workflow, the Jastrow factor is determined (typically by optimisation, such as with Variational Monte Carlo) for a specific system.


In this work, we consider the Drummond-Towler-Needs (DTN) form\cite{drummondJastrow,lopezriosFramework2012}
\begin{equation}
    \label{eq:dtn-jastrow}
    J_\mathrm{DTN}(L_{ee}, L_{en}, L_{een}) = \sum_{i<j}^Nv(r_{ij}) + \sum_i^N\sum_I^{N_A}\chi(r_{iI})
      + \sum_{i<j}^N\sum_I^{N_A}f(r_{ij}, r_{iI}, r_{jI}),
\end{equation}
with
\begin{equation}
    \label{eq:dtn-jastrow-ee}
    v(r_{ij})    = t(r_{ij},L_{ee})
                    \sum_{k} a_k r_{ij}^k ,
\end{equation}
\begin{equation}
    \label{eq:dtn-jastrow-en}
    \chi(r_{iI}) = t(r_{iI},L_{en})
    \sum_{k} b_k r_{iI}^k ,
\end{equation}
\begin{equation}
    \label{eq:dtn-jastrow-een}
    f(r_{ij}, r_{i}, r_{j}) = \,t(r_{iI},L_{een}) t(r_{jI},L_{een})
    \sum_{k,l,m} c_{klm}
    r_{ij}^k r_{iI}^l r_{jI}^m ,
\end{equation}
representing the e-e, e-n and e-e-n terms, respectively, and $t(r,L)$ is a cutoff function, such as
\begin{equation}
    t(r,L)=(1-r/L)^3\Theta(r-L).
\end{equation}

We then optimise\cite{hauptOptimizing2023} the parameters $\{a_k\}$, $\{b_k\}$, and $\{c_{klm}\}$, subject to the exact e-e and e-n cusp conditions (the latter imposed via the $\Lambda$ function described in \cite{hauptOptimizing2023}), by minimising the ``variance of the reference'',\cite{handyMinimization1971}
\begin{equation}
  \label{eq:var_eref}
    \sigma_\mathrm{ref}^2 = \bra{\Phi}|\HTC-E_\mathrm{ref}|^2\ket{\Phi},
\end{equation}
where $E_\mathrm{ref}$ is the reference energy and $\ket{\Phi}$ is the Hartree-Fock determinant. In practice, previous studies have used 8, 8, and 26 parameters in the e-e, e-n, and e-e-n parameters, respectively, to be optimised for atomic systems, and this number is only increased for multiatomic systems.

These flexible Jastrow factors have already proven to be successful in TC calculations in previous studies with cutoff values $L_{ee}=4.5, L_{en}=4$ and $L_{een}=4$.\cite{hauptOptimizing2023,christlmaierXTC2023,simulaEcps2025,filipDeterministic2025,filip_2ndrow,hauptMultireferenceTC2025}



\subsubsection{Simplified Jastrow Factors}
\label{sec:simple-jastrows}

On the extreme end, we might consider what we call ``minimal'' Jastrow factors, that is, those that resolve cusp conditions but do not include optimisable parameters. For example, the minimal Jastrow factor with e-e, e-n and e-e-n terms may be written
\cite{fournaisSharp2005,fournaisNonIsotropic2007}
\begin{equation}
    \label{eq:fournais-full}
    J_\mathrm{minimal} = \frac 12\sum_{i<j}r_{ij}t(r_{ij}, L_{ee}) -\sum_I\sum_i Z_Ir_{iI}t(r_{iI}, L_{en}) + \frac{2-\pi}{6\pi}\sum_I\sum_{i<j}Z_I\bm r_{iI}\cdot \bm r_{jI}\ln(r_{iI}^2+r_{jI}^2)t(r_{ijI}, L_{een}),
\end{equation}
where $t$ is a cutoff function as before. This form of the Jastrow factor, however, must either have such small cutoff lengths so as to not have a major effect on the total energies\cite{hauptDissertation2025} or have uncontrolled errors, with energies below the exact result.\cite{hauptDissertation2025,szenesStriking2024}

Instead, we wish to simplify the DTN Jastrow factors presented in section \ref{sec:flexible-jastrows} while keeping their flexibility. We propose breaking the system into modules, fixing the Jastrow components for those modules, and using them for the combined system. The most natural way of doing this for molecules is to consider each atom individually and reusing the Jastrow factors for the atoms as fixed parameters in the molecule. We propose reducing the cutoffs so that the terms are more local around the atom, and reoptimising the e-e term in the molecule (as this term is not inherently tied to a nucleus as the e-n or e-e-n terms are). For this, we use $L_{ee}=4.5$ a$_0$, as in section \ref{sec:flexible-jastrows}, and $L_{en}=1$ a$_0$, $L_{een}=2$ a$_0$. That is, given a set of Jastrow factor terms for atoms $\{J_\mathrm{atom}^{(ee)},J_\mathrm{atom}^{(en)},J_\mathrm{atom}^{(een)}\}$, we construct the Jastrow factor for the molecule as
\begin{equation}
    J_\mathrm{mod} = \sum_{i<j}^Nv(r_{ij}) + \sum_\mathrm{atom} (J_\mathrm{atom}^{(en)}+J_\mathrm{atom}^{(een)}),
\end{equation}
where $v(r)$ is defined in equation \ref{eq:dtn-jastrow-ee}, and contains the only parameters to be optimised using equation \ref{eq:var_eref} as the objective function.

For comparison, we also consider the full optimisation procedure to obtain $J_\mathrm{DTN}$ as described in section \ref{sec:flexible-jastrows}, as well as an e-e-only Jastrow factor,
\begin{equation}
J_{ee} = \sum_{i<j}^Nv(r_{ij}),
\end{equation}
so that our modular Jastrow factor has the same number of optimisable parameters. We refer to the Jastrow factor obtained by optimising the atoms and not reoptimising any terms (instead taking the e-e term of the heaviest atom) as $J_\mathrm{mod}|_\mathrm{no-opt}$.

\subsubsection{Comparison of Jastrow factors}
\label{sec:comparison-jastrows}

\begin{figure}[htbp]
    \centering
    \includegraphics[width=0.45\textwidth]{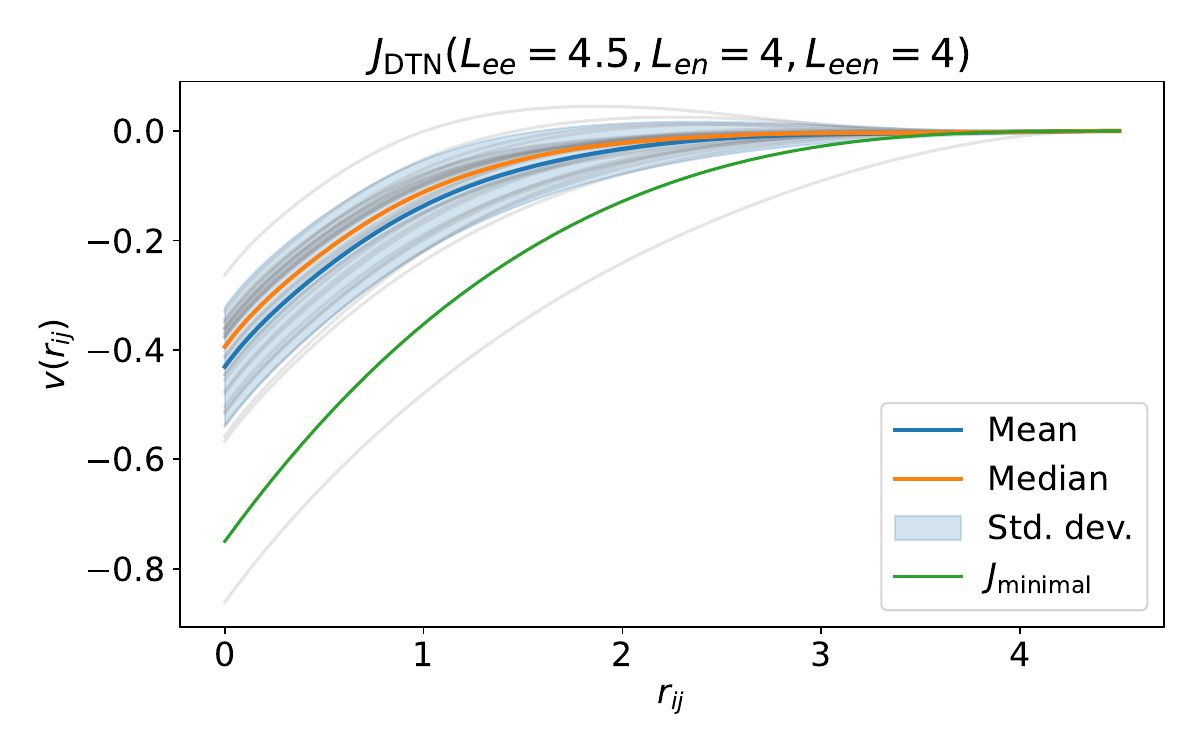}
    \includegraphics[width=0.45\textwidth]{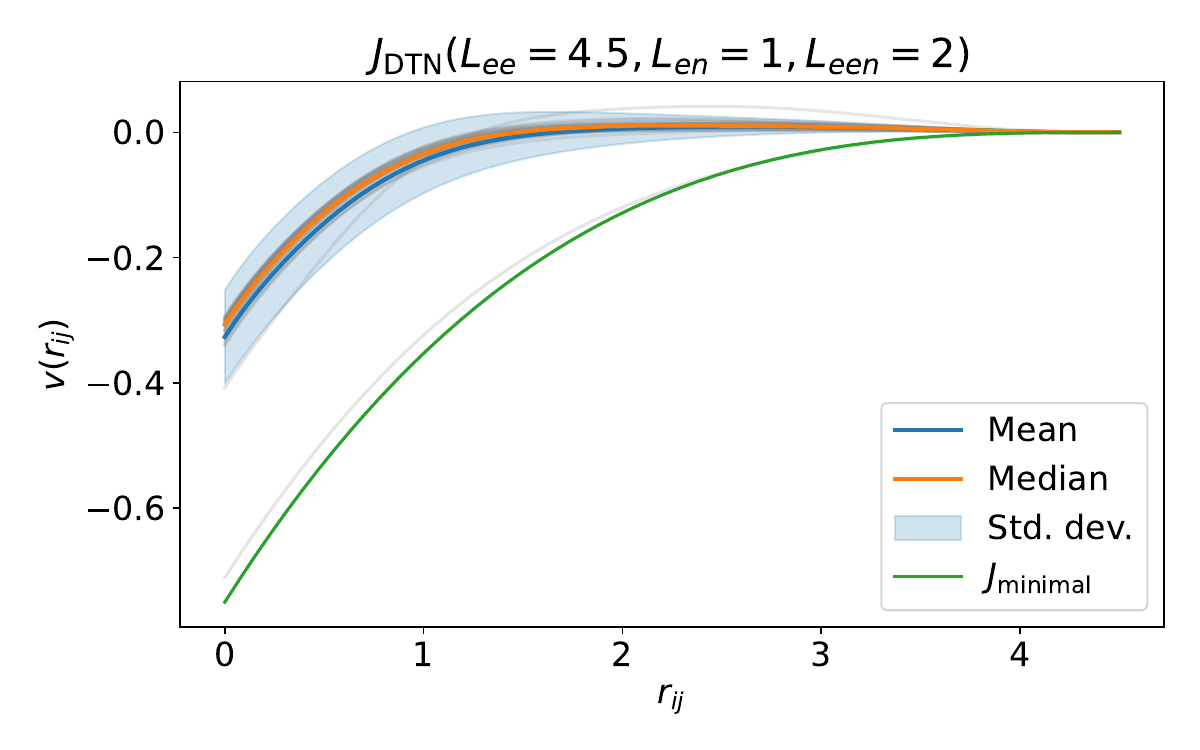}\\
    \includegraphics[width=0.45\textwidth]{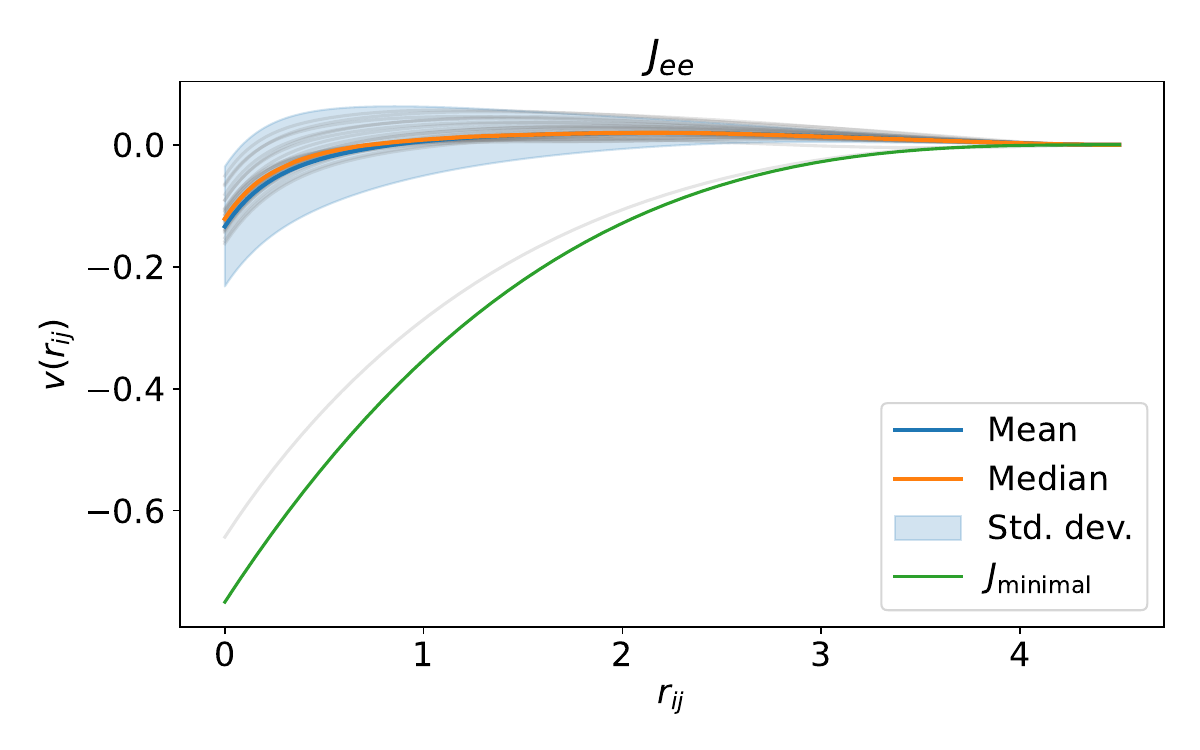}
    \includegraphics[width=0.45\textwidth]{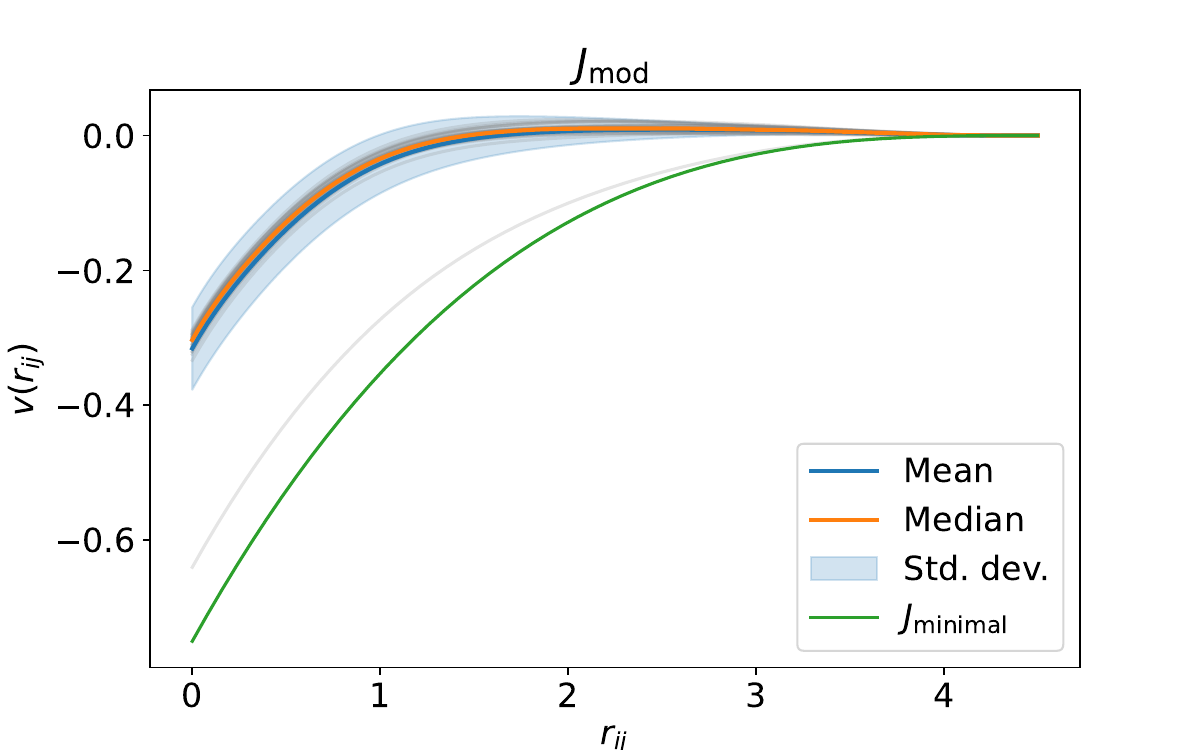}
    \caption{Clockwise from top left: the e-e terms for the $J_\mathrm{DTN}(L_{ee}=4.5, L_{en}=4, L_{een}=4)$, $J_\mathrm{DTN}(L_{ee}=4.5, L_{en}=1, L_{een}=2)$ (i.e. all terms optimised), $J_\mathrm{mod}$ (e-e term reoptimised) and $J_{ee}$ (e-e term only) Jastrow factors for the HEAT set of molecules with the \avtz basis set. Included in each subplot is also the minimal e-e Jastrow factor $\frac 12 r_{12}(1-r_{12}/L_{ee})$ where $L_{ee}=4.5$ bohr. The significant outlier in each plot is the Jastrow factor of H$_2$.
    }
    \label{fig:ee-jastrow-spread}
\end{figure}

If we plot the e-e component $v(r_{ij})$ of the Jastrow factors for the HEAT set of molecules\cite{tajtiHEAT2004,bombleHighaccuracy2006_HEAT2,hardingHighaccuracy2008_HEAT3,thorpeHighaccuracy2019_HEAT4}, we find that the spread is substantially higher for the full DTN Jastrow factor with $(L_{ee}, L_{en}, L_{een})=(4.5,4,4)$ compared to using the smaller cutoffs $(L_{ee}, L_{en}, L_{een})=(4.5,1,2)$, especially with the reoptimised modular Jastrow factor. The e-e-only Jastrow factor $J_{ee}$ also has reduced spread but not as drastically as the modular Jastrow factor. This is shown in figure \ref{fig:ee-jastrow-spread}. From these plots, we see that the minimal Jastrow factor fails to universally capture the behaviour of the optimised terms.


Total and reference energies for N and N$_2$ are included in table \ref{tbl:n_n2_jastrow_energies} for the choices of Jastrow factors, along with essentially-exact benchmark values calculated with HEAT.\cite{tajtiHEAT2004,bombleHighaccuracy2006_HEAT2,hardingHighaccuracy2008_HEAT3,thorpeHighaccuracy2019_HEAT4} Note that due to the minimal Jastrow factor's overestimate for the correlation hole, as shown in figure \ref{fig:ee-jastrow-spread}, energy estimates with this Jastrow factor fall far below the exact result. xTC-CCSD(T) largely corrects this, with a large positive correlation energy, but the energy still falls below the exact result. In contrast, the xTC-CCSD(T) energies for all the Jastrow factors involving optimisable parameters are above the exact result. This shows that the use of flexible Jastrow functions, optimised using the reference-energy variance minimisation procedure of \cite{hauptOptimizing2023}, generally yields energies above the exact value.

\begin{table}[htbp]
    \centering

    \begin{tabular}{c|ccccc|c}
        \multicolumn{6}{c}{\textbf{Single-determinant TC Reference Energies}} \\
        System & Non-TC & $J_\mathrm{minimal}|_{ee}$ & $J_{ee}$ & $J_\mathrm{DTN}(4.5, 1, 2)$ & $J_\mathrm{DTN}(4.5, 4, 4)$ & HEAT \\
        \hline
        N & -54.3976 & -57.2711 & -54.4585  & -54.4834 & -54.5178  & -54.5897 \\
        N$_2$ & -108.9846 & -119.7395 & -109.1294 & -109.1858 & -109.2574 & -109.5435 \\
        \hline
        Atomisation (mHa) & 189.4 & 5197.3 & 212.4 & 219.0 & 221.8 & 364.1
    \end{tabular}

    \vspace{5mm}

    \begin{tabular}{c|ccccc|c}
        \multicolumn{6}{c}{\textbf{xTC-CCSD(T) Energies}} \\
        System & Non-TC & $J_\mathrm{minimal}|_{ee}$ & $J_{ee}$ & $J_\mathrm{DTN}(4.5, 1, 2)$ & $J_\mathrm{DTN}(4.5, 4, 4)$ & HEAT \\
        \hline
        N & -54.5275 & -54.7690 & -54.5668 & -54.5842 & -54.5852 & -54.5897 \\
        N$_2$ & -109.4114 & -110.3230 & -109.4903 & -109.5316 & -109.5334 & -109.5435 \\
        \hline
        Atomisation (mHa) & 356.4 & 785.0 & 356.7 & 363.2 & 363.0 & 364.1
    \end{tabular}
    \caption{TC Hartree-Fock (reference) energies $\bra{\Phi_\mathrm{HF}}\HTC\ket{\Phi_\mathrm{HF}}$ (top) and xTC-CCSD(T) energies (bottom) for N and N$_2$ at the equilibrium geometry, using the various choices of Jastrow factors discussed in the text, in the \avtz basis. Non-transcorrelated energies at the level of HF and CCSD(T) are provided for ease of reference. $J_\mathrm{minimal}|_{ee}$ refers to the e-e-only minimal Jastrow factor $\frac 12 \sum_{i<j}r_{ij}t(r_{ij}, L_{ee})$. Values obtained using the high-accuracy method HEAT\cite{tajtiHEAT2004,bombleHighaccuracy2006_HEAT2,hardingHighaccuracy2008_HEAT3,thorpeHighaccuracy2019_HEAT4} are included as a benchmark.
    \label{tbl:n_n2_jastrow_energies} It is clear that the minimal Jastrow factor produces massive non-variationality at the level of the reference energies, which is reduced, but not eliminated, by the post-HF xTC-CCSD(T) method. On the other hand, an e-e-only Jastrow factor with optimisable terms eliminates the undesirable non-variational behaviour. The more flexible DTN forms, with e-n and e-e-n terms, yield improved total and atomisation energies, implying overall improvement as the flexibility of the Jastrow factor is increased. Relatively small dependence of the atomisation energies on the cutoff lengths of the e-n and e-e-n terms can also be noted.}
\end{table}

\begin{figure}[htbp]
    \centering
    \includegraphics[width=0.45\textwidth]{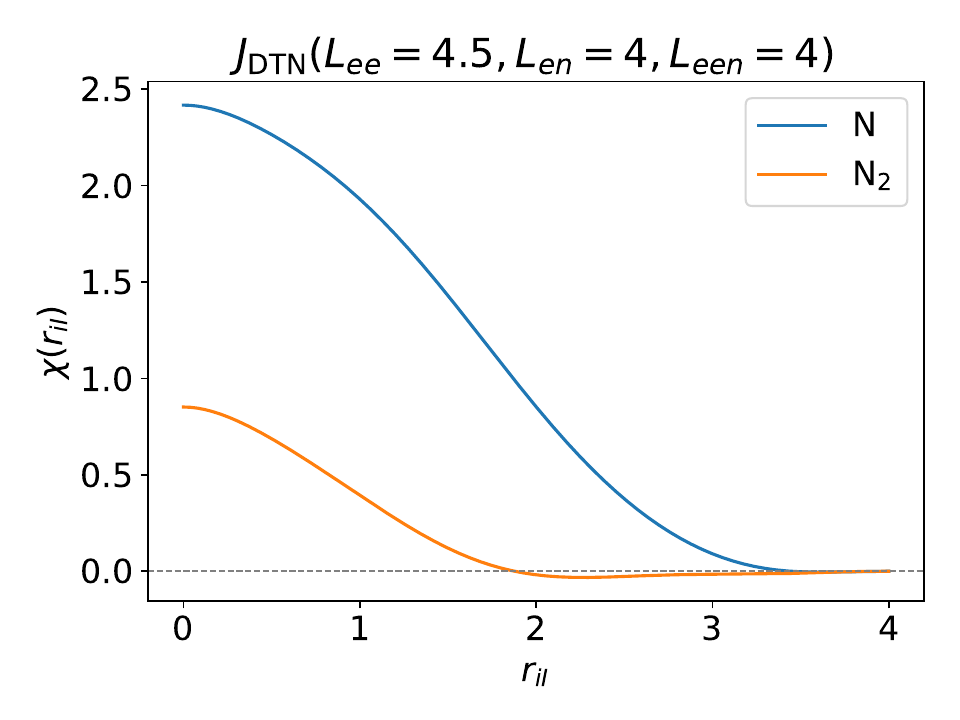}
    \includegraphics[width=0.45\textwidth]{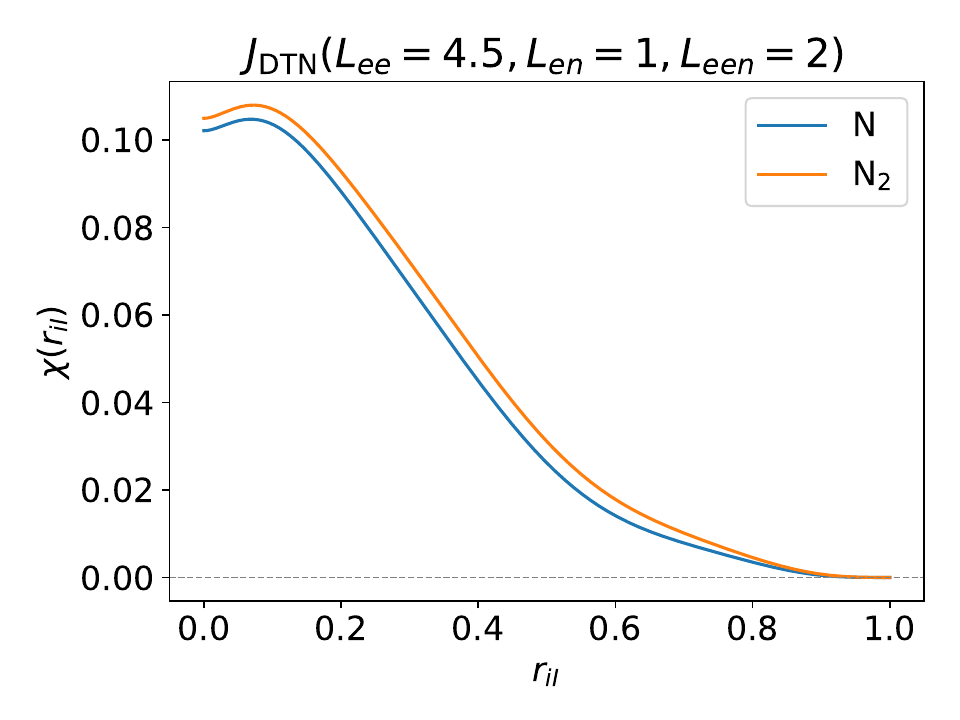}
    \caption{e-n Jastrow factor terms for the cutoff combinations $L_{ee}=4.5, L_{en}=4, L_{een}=4$ and $L_{ee}=4.5, L_{en}=1, L_{een}=2$ for N$_2$ and the N atom. From these plots, we see that the shorter cutoff lengths provide forms where the atomic and molecular Jastrow factors are more similar, due to the more local character of these Jastrow factors.
    }
    \label{fig:en-N-N2}
\end{figure}

To further motivate the reduction of the cutoff lengths for the modular Jastrow factor, figure \ref{fig:en-N-N2} shows the e-n component of the $J_\mathrm{DTN}(L_{ee}=4.5, L_{en}=4, L_{een}=4)$ and $J_\mathrm{DTN}(L_{ee}=4.5, L_{en}=1, L_{een}=2)$ Jastrow factors for N$_2$ and the N atom. It is clear from this plot that the latter case has an atomic Jastrow factor better suited for the molecule. We therefore continue to employ these cutoff lengths for the modular Jastrow factor form.

\begin{figure}[htbp]
    \centering
    \includegraphics[width=0.8\textwidth]{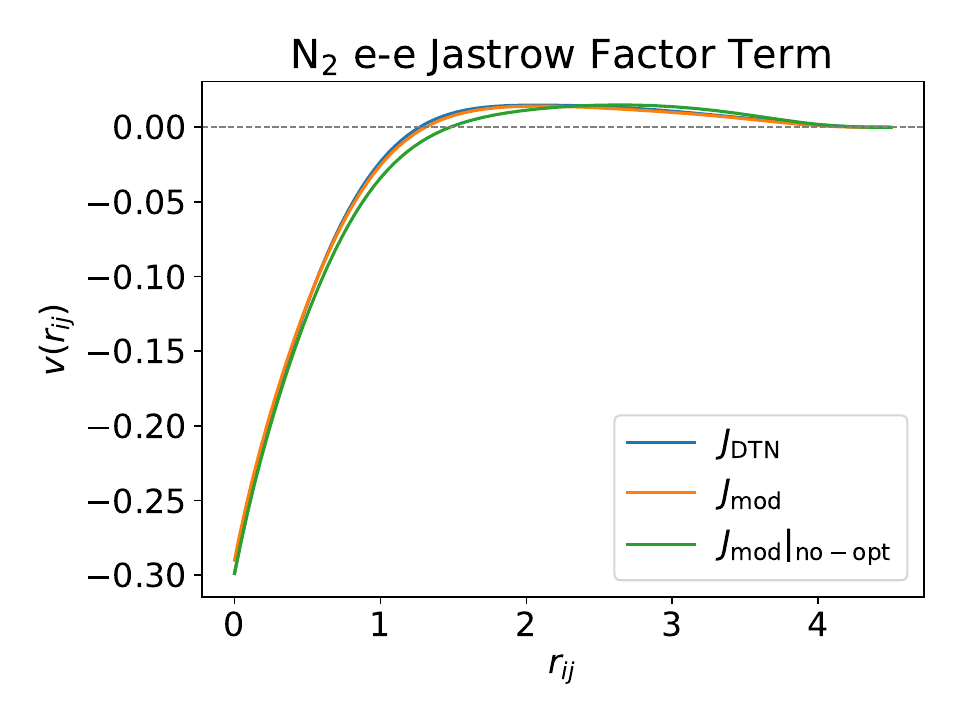}
    \caption{e-e Jastrow factor terms for N$_2$ using $J_\mathrm{DTN}(L_{ee}=4.5, L_{en}=1, L_{een}=2)$ (full optimisation), $J_\mathrm{mod}$ (optimising only e-e component from the atom) and $J_\mathrm{mod}|_\mathrm{no-opt}$ (no further optimisation, simply use the atom's Jastrow factor), with the \avtz basis set. While all three are similar to each other, we see that by optimising the atom's Jastrow factor, we return to a form particularly similar to the fully-optimised form.
    }
    \label{fig:ee-compare-full-no-reopt}
\end{figure}

We also compare how the e-e component behaves for the modular Jastrow factor when optimised compared to no optimisation. Note that a similar approach has been used in previous studies\cite{cohenSimilarity2019}, where fixed Jastrow factors such as SM17\cite{schmidtCorrelated1990} were used for both the ion and the atom when calculating the ionisation potential. Comparing atomic and molecular Jastrow factors, figure \ref{fig:ee-compare-full-no-reopt} shows that while similar to the molecule's e-e Jastrow term, reoptimising this term from the atom brings it closer to the form obtained from optimising all terms, possibly indicating an improved Jastrow factor parametrisation.

\begin{figure}[htbp]
    \centering
    \includegraphics[width=0.8\textwidth]{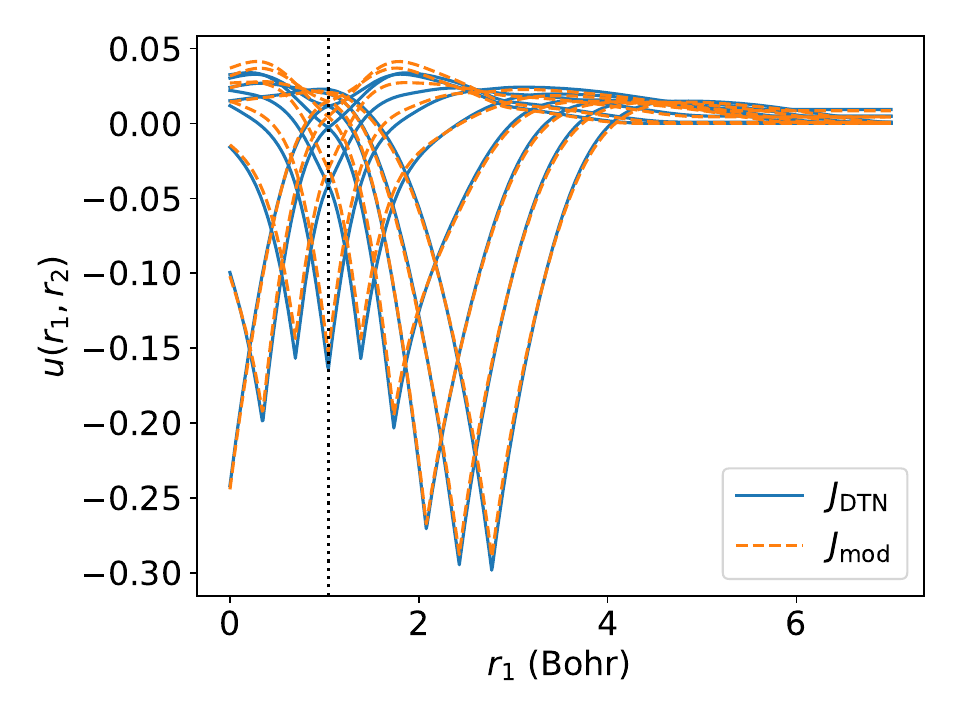}
    \caption{$u(r_1,r_2)$ (see equation \ref{eq:ur1r2def}) for the fully-optimised $J_\mathrm{DTN}(L_{ee}=4.5, L_{en}=1, L_{een}=2)$ (blue solid curves) and e-e-reoptimised $J_\mathrm{mod}$ (orange dashed curves) Jastrow factors. Note that by the molecule's symmetry, with nitrogen nuclei located at $\pm 1.04$ bohr, these Jastrow factors are symmetric about zero, so we plot only for $r_1\ge 0$ and indicate the nucleus' position with a dotted vertical line. The location of the fixed electron $r_2$ corresponds to the prominent cusps which deepen as the electrons are further from the nucleus.
    }
    \label{fig:ur1r2jastrow}
\end{figure}

As a final comparison, we note that these Jastrow factors may be rewritten
\begin{equation}
    \label{eq:ur1r2jastrow}
    J=\sum_{ij}u(\bm r_i,\bm r_j)
\end{equation}
where
\begin{equation}
    \label{eq:ur1r2def}
    u(\bm r_i,\bm r_j) = v(r_{ij})+\frac{1}{N-1}\sum_I[\chi(r_{iI})+\chi(r_{jI})]
    + \sum_I f(r_{ij}, r_{iI}, r_{jI})
\end{equation}
and $N$ is the number of electrons. To recast this in a form that is easy to interpret, we plot $u(r_1,r_2)$ for a special case: the nitrogen dimer, with one of the electron positions $r_2$ fixed at various locations and the other electron $r_1$ scanned along the bond axis. Such a plot is depicted in figure \ref{fig:ur1r2jastrow}. From this figure, we see that the fully-optimised and the modular e-e-reoptimised Jastrow factors have similar shapes. Note that unlike figures \ref{fig:ee-jastrow-spread}, \ref{fig:en-N-N2} and \ref{fig:ee-compare-full-no-reopt}, this plot contains all terms: e-e, e-n and e-e-n.

\section{Results and Discussion}

\subsection{Computational Details}

In the following discussion, we compare the choices of Jastrow factors when calculating atomisation energies, ionisation potentials and calculating the binding curve of carbon monoxide (CO). In these calculations, \molpro\cite{wernerMOLPRO1,wernerMolpro2_2020,wernerMolpro2012_3} was used to generate the orbitals, \casino\cite{needsCASINO2020} was used to optimise the Jastrow factors, \pytchint\cite{tchint} was used to integrate TC integrals over a Becke grid\cite{beckeMulticenter1988} generated by \pyscf\cite{sunPySCF2018,sunRecent2020}, coupled cluster calculations such as biorthogonal xTC-CCSD(T) (that is, coupled cluster with singles and double excitations and perturbative triples performed with xTC integrals)\cite{katsOrbital2024} were carried out with \elemcojl\cite{katsElemCojl2025} and xTC-FCIQMC was performed with \neci\cite{gutherNECI2020}.

\subsection{Atomisation Energies}
\label{sec:heat-atomisation}

\begin{figure}[htbp]
    \centering
    \includegraphics[width=0.8\textwidth]{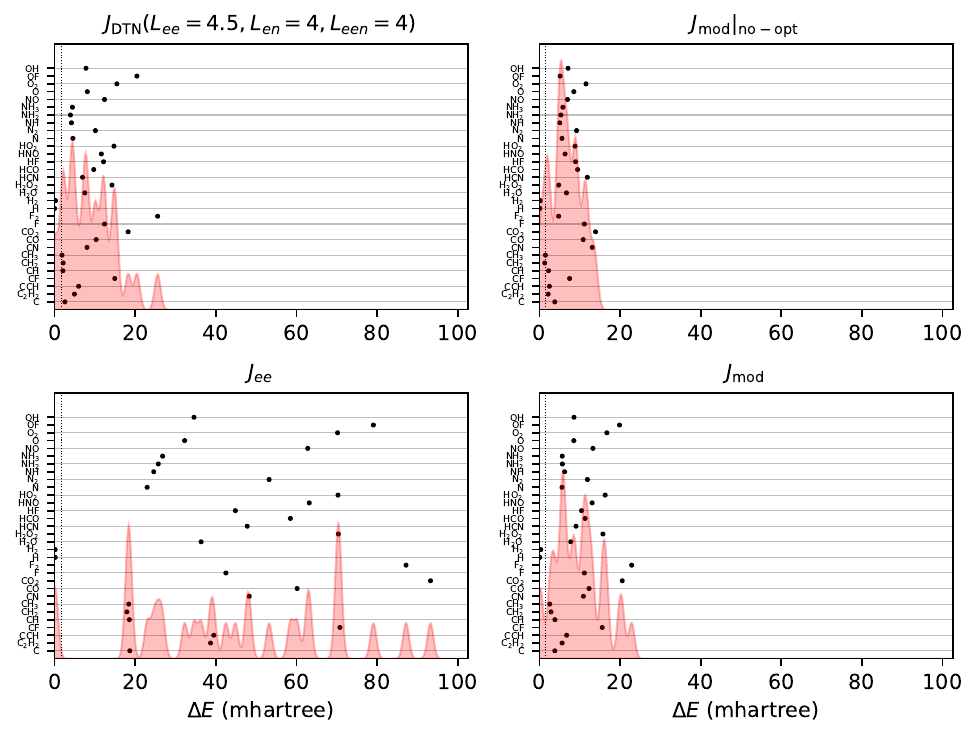}
    \caption{
    Errors in the total xTC-CCSD(T) energies for the HEAT set of molecules and atoms with the \avtz basis set and (clockwise from top-left): the $J_\mathrm{DTN}(L_{ee}=4.5, L_{en}=4, L_{een}=4)$, $J_\mathrm{mod}|_\mathrm{no-opt}$, $J_\mathrm{mod}$ and $J_{ee}$ Jastrow factors. Dotted lines indicate chemical accuracy (1.6 mHa). The red shaded areas are a visual aid only: they correspond to a sum of gaussians centred on the data points with the width of the Gaussians chosen such that equidistantly distributed Gaussians would be contained to 95\% in the corresponding segment.
    }
    \label{fig:heat-energy}
\end{figure}

We present the total energies of the 31 atoms and molecules in the HEAT benchmark set\cite{tajtiHEAT2004,bombleHighaccuracy2006_HEAT2,hardingHighaccuracy2008_HEAT3,thorpeHighaccuracy2019_HEAT4} for the \avtz basis set\cite{kendallElectron1992} using xTC-CCSD(T) and the corresponding atomisation energies. The errors in the total energies compared to HEAT for the $J_\mathrm{DTN}(L_{ee}=4.5, L_{en}=4, L_{een}=4)$ (which we use as a benchmark, corresponding to the approach used in previous studies), $J_\mathrm{mod}$, $J_\mathrm{mod}|_\mathrm{no-opt}$ and $J_{ee}$ Jastrow factors are presented in figure \ref{fig:heat-energy}. We note from this figure that the e-e-only Jastrow factor $J_{ee}$ is the least accurate, indicating the need for terms involving the nuclei. The other Jastrow forms perform similarly to each other.

\begin{figure}[htbp]
    \centering
    \includegraphics[width=0.8\textwidth]{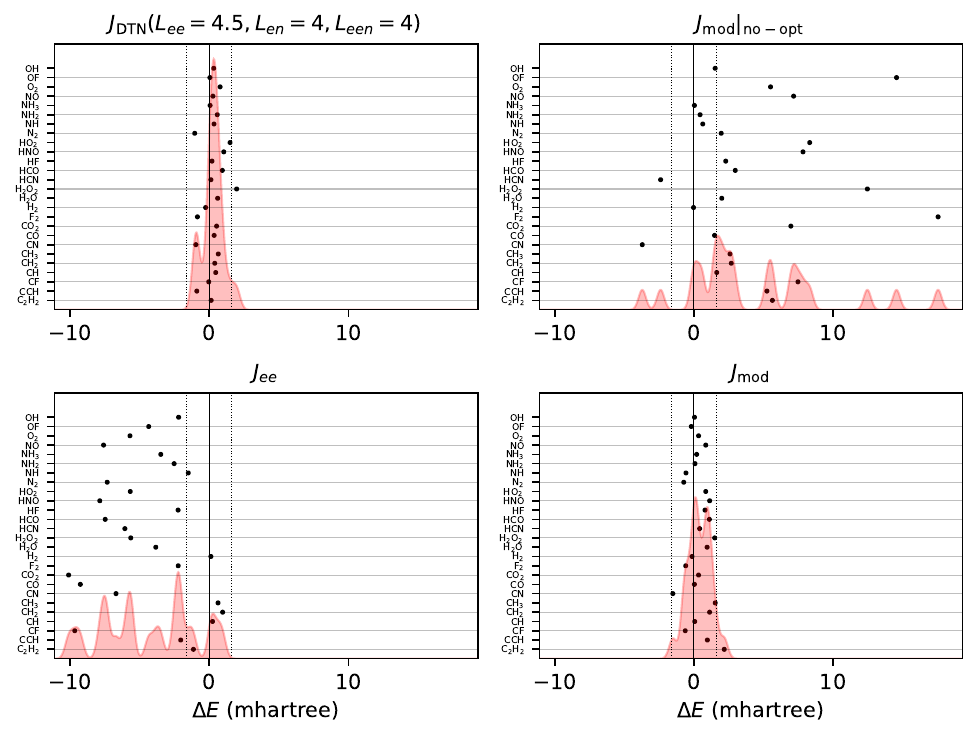}
    \caption{Errors in the xTC-CCSD(T) atomisation energies for the HEAT set of molecules and atoms with the \avtz basis set and (clockwise from top-left): the $J_\mathrm{DTN}(L_{ee}=4.5, L_{en}=4, L_{een}=4)$, $J_\mathrm{mod}|_\mathrm{no-opt}$, $J_\mathrm{mod}$ and $J_{ee}$ Jastrow factors. Dotted lines indicate chemical accuracy (1.6 mHa). The red shaded areas are a visual aid only: they correspond to a sum of gaussians centred on the data points with the width of the Gaussians chosen such that equidistantly distributed Gaussians would be contained to 95\% in the corresponding segment.
    }
    \label{fig:heat-atomisation}
\end{figure}

The corresponding atomisation energies for these Jastrow factors are shown in figure \ref{fig:heat-atomisation}. As before, $J_{ee}$ does not perform as well as the others. However, also noteworthy is that while the total energies for $J_\mathrm{mod}$ and $J_\mathrm{mod}|_\mathrm{no-opt}$ (as shown in figure \ref{fig:heat-energy}) are similar and above the exact result, only the $J_\mathrm{DTN}$ and $J_\mathrm{mod}$ Jastrow factors are largely within chemical accuracy in the atomisation energies. This implies that favourable error cancellation hinges on careful optimisation of the e-e term, as small changes as shown in figure \ref{fig:ee-compare-full-no-reopt} can lead to a several mHa error in the atomisation energies.

\subsection{Ionisation Potentials}

To show the flexibility of the modular Jastrow factor method to also handle charged systems, we present the ionisation potentials of the first row atoms using xTC-CCSD(T) for \avdz and \avtz and compare against standard (non-TC) CCSD(T). In this context, the ``module'' for the ion is simply the atom, so $J_\mathrm{mod}$ shares e-n and e-e-n terms between atom and ion, but the e-e term is reoptimised in the ion.

\begin{table}[htbp]
    \centering
    \begin{tabular}{clrrrrrrrr}
        & Method & Li & Be & B & C & N & O & F & Ne  \\
        \midrule
        \multirow{4}{*}{\rotatebox[origin=c]{90}{avdz}} & \multicolumn{1}{|l}{Non-TC} & 196.43 & 341.24 & 297.51 & 406.78 & 527.71 & 483.31 & 629.09 & 787.32 \\
        & \multicolumn{1}{|l}{$J_\mathrm{DTN}$} & 197.04 & 344.11 & 301.27 & 411.32 & 533.44 & 496.26 & 641.42 & 798.97 \\
        & \multicolumn{1}{|l}{$J_\mathrm{mod}$} & 197.77 & 344.99 & 302.22 & 412.41 & 534.13 & 494.58 & 640.75 & 798.60 \\
        & \multicolumn{1}{|l}{$J_\mathrm{mod}|_\mathrm{no-opt}$} & 198.45 & 343.94 & 300.92 & 411.12 & 533.59 & 495.01 & 641.46 & 801.36 \\
        \midrule
        \multirow{4}{*}{\rotatebox[origin=c]{90}{avtz}} & \multicolumn{1}{|l}{Non-TC} & 196.74 & 341.22 &  302.55  & 411.68 & 532.52 & 493.58 & 634.86 & 789.02 \\
        & \multicolumn{1}{|l}{$J_\mathrm{DTN}$} & 197.87 & 342.53 &  304.19  & 413.62 & 534.74 & 500.42 & 641.89 & 796.09 \\
        & \multicolumn{1}{|l}{$J_\mathrm{mod}$} & 198.04 & 342.62 &  304.39  & 413.68 & 534.94 & 499.42 & 640.56 & 794.54 \\
        & \multicolumn{1}{|l}{$J_\mathrm{mod}|_\mathrm{no-opt}$} & 198.47 & 342.92 &  304.58  & 414.07 & 535.32 & 499.38 & 641.43 & 796.50 \\
        \midrule
        & Experiment & 198.15 & 342.58 & 304.99 & 413.97 & 534.60 & 500.50 & 641.10 & 794.50
    \end{tabular}
    \caption{Ionisation potentials for the first row atoms in mHa. Experimental values\cite{chakravortyGroundstate1993} are also provided. Here the cutoffs are $L_{ee}=4.5,L_{en}=1,L_{een}=2$ bohr and as before, for both $J_\mathrm{mod}$ and $J_\mathrm{DTN}$; $J_\mathrm{mod}|_\mathrm{no-opt}$ refers to $J_\mathrm{mod}$ with the e-e terms also fixed, i.e. no optimisation is performed on the ion. The abbreviations ``avdz'' and ``avtz'' refer to basis sets \avdz and \avtz respectively.
        }
    \label{tab:ip-results-raw}
\end{table}

Ionisation potentials are provided in table \ref{tab:ip-results-raw}. The error as compared to experiment \cite{chakravortyGroundstate1993} is shown in table \ref{tab:ip-results}. The tables show that, particularly for heavier atoms, transcorrelated methods far outperform the standard CCSD(T). Moreover, $J_\mathrm{DTN}$ and $J_\mathrm{mod}$ are entirely within chemical accuracy (defined here as 1.6 mHa) at \avtz, with the no-optimisation method also performing well. The fact that $J_\mathrm{mod}$ slightly outperforms $J_\mathrm{DTN}$ despite being a simpler, less flexible form, might indicate favourable error cancellation from the ion's and atom's shared e-n and e-e-n terms. Since optimising e-e results in reduced errors, this indicates that optimising this term is critical for favourable error differences, as in section \ref{sec:heat-atomisation}.

\begin{table}[htbp]
    \centering
    \begin{tabular}{clrrrrrrrr|r}
        & Method & Li & Be & B & C & N & O & F & Ne & MAE \\
        \midrule
        \multirow{4}{*}{\rotatebox[origin=c]{90}{avdz}} & \multicolumn{1}{|l}{Non-TC} & -1.72 & -1.34 & -7.48 & -7.19 & -6.89 & -17.19 & -12.01 & -7.18 & 7.625 \\
        & \multicolumn{1}{|l}{$J_\mathrm{DTN}$} & -1.11 & 1.53 & -3.72 & -2.65 & -1.16 & -4.24 & 0.32 & 4.47 & 2.400 \\
        & \multicolumn{1}{|l}{$J_\mathrm{mod}$} & -0.38 & 2.41 & -2.77 & -1.56 & -0.47 & -5.92 & -0.35 & 4.10 & 2.245 \\
        & \multicolumn{1}{|l}{$J_\mathrm{mod}|_\mathrm{no-opt}$} & 0.30 & 1.36 & -4.07 & -2.85 & -1.01 & -5.49 & 0.36 & 6.86 & 2.788 \\
        \midrule
        \multirow{4}{*}{\rotatebox[origin=c]{90}{avtz}} & \multicolumn{1}{|l}{Non-TC} & -1.41 & -1.36 & -2.44 & -2.29 & -2.08 & -6.92 & -6.24 & -5.48 & 3.528 \\
        & \multicolumn{1}{|l}{$J_\mathrm{DTN}$} & -0.28 & -0.05 & -0.80 & -0.35 & 0.14 & -0.08 & 0.79 & 1.59 & 0.510 \\
        & \multicolumn{1}{|l}{$J_\mathrm{mod}$} & -0.11 & 0.04 & -0.60 & -0.29 & 0.34 & -1.08 & -0.54 & 0.04 & 0.380 \\
        & \multicolumn{1}{|l}{$J_\mathrm{mod}|_\mathrm{no-opt}$} & 0.32 & 0.34 & -0.41 & 0.10 & 0.72 & -1.12 & 0.33 & 2.00 & 0.667 \\
    \end{tabular}
    \caption{Error in the ionisation potentials for the presented atom and method, relative to experimental values calculated by experiment \cite{chakravortyGroundstate1993}, $E_{IP}-E_{exp}$. Values are in mHa. Here the cutoffs are $L_{ee}=4.5,L_{en}=1,L_{een}=2$ bohr and as before, for both $J_\mathrm{mod}$ and $J_\mathrm{DTN}$; $J_\mathrm{mod}|_\mathrm{no-opt}$ refers to $J_\mathrm{mod}$ with the e-e terms also fixed, i.e. no optimisation is performed on the ion. The abbreviations ``avdz'' and ``avtz'' refer to basis sets \avdz and \avtz respectively.}
    \label{tab:ip-results}
\end{table}

\subsection{Carbon Monoxide Binding Curve}

\begin{figure}[htbp]
    \centering
    \includegraphics[width=0.8\textwidth]{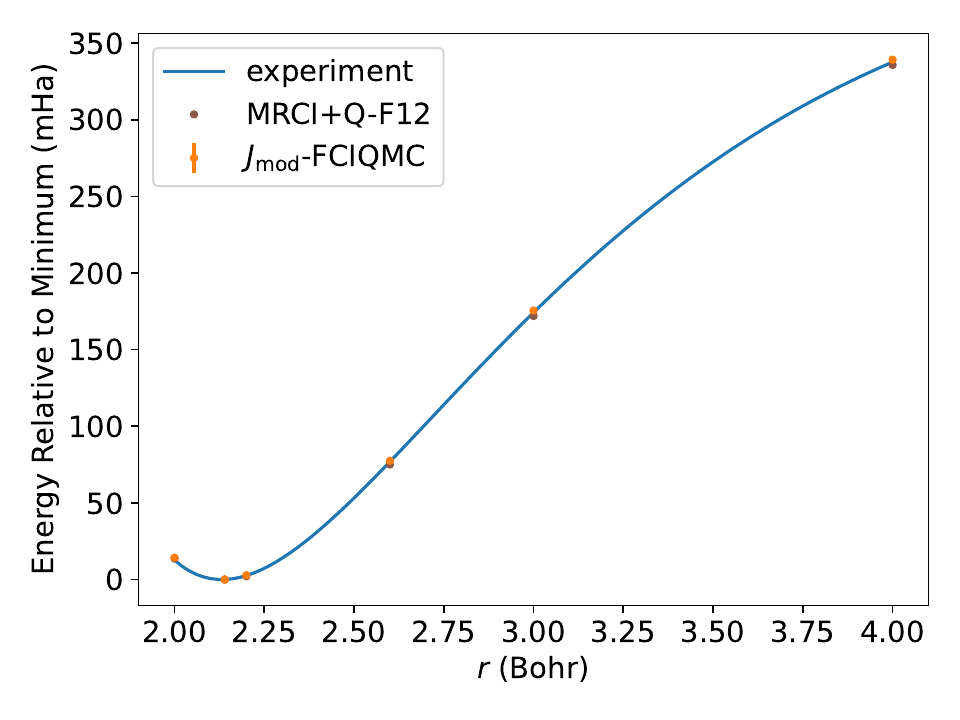}
    \includegraphics[width=0.8\textwidth]{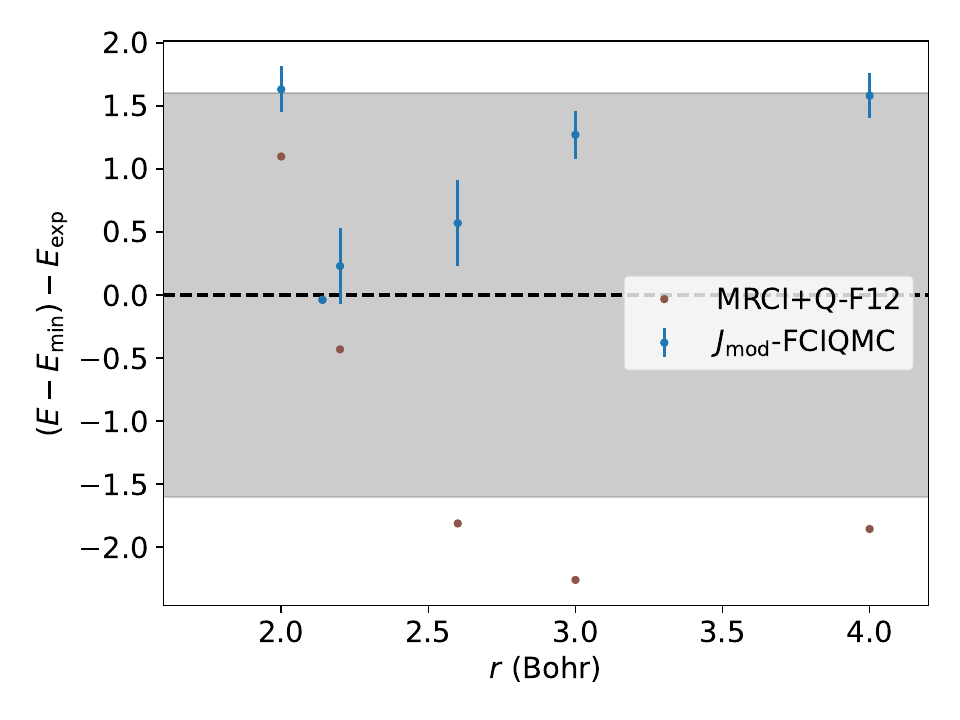}
    \caption{
    Energy of the carbon monoxide molecule (top) and that value relative to experiment\cite{CoxonDirect2004} (bottom) as a function of internuclear distance $r$. Curves have been normalised such that the equilibrium ($r=2.14$ bohr) energy value is zero.
    Calculations are performed with the \avtz basis set. xTC-FCIQMC calculations are provided for the modular Jastrow factor, and MRCI+Q-F12 is provided as reference. In the bottom panel, the dashed line at zero represents the experimental value and the shaded region $\pm 1.6$ mHa. As seen, the xTC-FCIQMC binding curve in the stretched region beyond 2.5 bohr outperforms MRCI+Q-F12 for this system.
        }
    \label{fig:co-binding}
\end{figure}

The binding curve of the carbon monoxide molecule in the \avtz basis is calculated using this workflow. The orbitals and reference wavefunction used for the Jastrow optimisation are calculated using unrestricted Hartree-Fock theory, avoiding the need for multireference optimisation\cite{hauptMultireferenceTC2025}. The electron-nuclear and electron-electron-nuclear terms of the C and O Jastrow factor are taken from atomic optimised parameters and reused in the molecule, and only the e-e term (with 8 parameters) is optimised at each geometry. Final energies presented in figure \ref{fig:co-binding} are from xTC-FCIQMC calculations, which were also performed in the same UHF basis. The xTC-FCIQMC calculations were performed with up to $3\times 10^8$ walkers, and the walker-extrapolation method was used\cite{hauptOptimizing2023} to obtain the infinite-walker limit. We find that this workflow performs remarkably well, within chemical accuracy of the experimental binding curve over a wide range of distances, and out-performing MRCI+Q-F12 in the same basis in the intermediate stretched region. The fact that these minimally optimised Jastrow factors work well over a range of distances bodes well for the use of this method to generate potential energy surfaces in other challenging, multiple bond-breaking, systems.

\section{Conclusion and Outlook}

We have presented a new approach to constructing Jastrow factors for use in the transcorrelated method. In particular, we have shown that we can partially reuse Jastrow factors from other systems, notably those of atoms. This method substantially reduces the number of terms needed for optimisation, simplifying the workflow, and even improving accuracy. Important future directions include building a Jastrow factor database for each atom and basis set, as well as combining the method with deterministic optimisation\cite{filipDeterministic2025} for added reliability and reproducibility, and using pseudopotentials\cite{simulaEcps2025} to scale to larger systems. Here we used Jastrow factors of similar parametrisation as previous works, but since we have shown that we need not optimise all parameters, it may be possible to construct more complex Jastrow factors to capture a greater deal of correlation and store these forms in a database to be recalled without additional complexity in the molecule. Furthermore, there may be systems where the individual ``modules'' need not be atoms; this is another potential future study. Finally, for the dissemination of the method and its usability, our software \pytchint will undergo further improvements before release so that it may be used easily with this approach.

\section{Acknowledgements}

The authors gratefully acknowledge the support of the Max Planck Society.

\printbibliography

\end{document}